\documentclass[12pt]{article}

\ifx\pdfoutput\undefined
\usepackage[dvips,bookmarks]{hyperref}
\else
\usepackage{hyperref}
\fi
\hypersetup{colorlinks=false,bookmarksopen,bookmarksnumbered,citecolor=blue,
   pdfstartview=FitH}

\usepackage{latexsym}

\usepackage{amssymb,amsfonts,amsmath}
\usepackage{graphicx} 
\usepackage{indentfirst}

 \usepackage{bbm}

\parskip=\medskipamount

\newcommand {\cL}{{\cal L}}

\newcommand {\cN}{{\cal N}}
\newcommand {\cO}{{\cal O}}

\newcommand {\cV}{{\cal V}}

\def\a{\alpha}

\def\b{\beta}

\def\d{\delta}
\def\e{\epsilon}

\def\j{\psi}

\def\l{\lambda}

\def\q{\theta}

\def\s{\sigma}

\def\x{\xi}
\def\z{\zeta}

\def\F{\Phi}
\def\J{\Psi}
\def\L{\Lambda}

\def\S{\Sigma}

\def\X{\Xi}

\def\rd{{\rm d}}
\def\ri{{\rm i}}
\def\re{{\rm e}}

\newcommand{\ad}{{\dot{\alpha}}}                           %new
\newcommand{\bd}{{\dot{\beta}}}                            %new
\newcommand{\ve}{\varepsilon}                            %new
\newcommand{\pa}{\partial}                           %new
\newcommand{\hf}{\frac12}
\newcommand{\vf}{\varphi}
\newcommand{\be}{\begin{equation}}
\newcommand{\ee}{\end{equation}}
\newcommand{\bea}{\begin{eqnarray}}
\newcommand{\eea}{\end{eqnarray}}

\newcommand{\ba}{\begin{array}}
\newcommand{\ea}{\end{array}}

\newcommand{\bm}[1]{\mbox{\boldmath$#1$}}

\def\double #1{#1{\hbox{\kern-2pt $#1$}}}

\newcommand{\bsubeq}{\begin{subequations}}
\newcommand{\esubeq}{\end{subequations}}

\begin{document}

\begin{titlepage}

\begin{flushright}
May 2011\\
\end{flushright}
\vspace{5mm}

\begin{center}
{\Large \bf  Goldstino superfields for spontaneously broken $\bm{\cN = 2}$ supersymmetry }
\end{center}

\begin{center}

{\large  
S. M. Kuzenko\footnote{kuzenko@cyllene.uwa.edu.au} and I. N. McArthur\footnote{{mcarthur@physics.uwa.edu.au}}
} \\
\vspace{5mm}

\footnotesize{
{\it School of Physics, M013, 
The University of Western Australia\\
35 Stirling Highway, Crawley W.A. 6009 Australia }}  
~\\

\vspace{2mm}

\end{center}
\vspace{5mm}

\begin{abstract}
\baselineskip=14pt
We examine spontaneously broken ${\cal N} = 2$ supersymmetry in four dimensions
and  associate a spinor superfield with each Goldstino via a finite supersymmetry transformation with parameters that are the Grassmann  coordinates of ${\cal N} = 2$ superspace. 
Making use of  a special choice of coset parametrization allows us to develop   
a version of nonlinearly realized ${\cal N}=2$ supersymmetry for which the associated Goldstino superfields 
are defined on harmonic superspace, 
thereby providing a natural mechanism for construction of a  Goldstino action.
The corresponding superfield Lagrangian is an $\cO (4)$ multiplet. 
This property is used to reformulate the Goldstino action in projective superspace 
and in conventional $\cN=2$ superspace.
We show how to generate matter couplings of the Goldstinos to supersymmetric matter  
using the $\cN=2$ harmonic, projective and full superspaces.
As a bi-product of our consideration, we also derive an $\cN=2$ chiral Goldstino action.
\end{abstract}
\vspace{1cm}

\vfill
\end{titlepage}

\tableofcontents{}
\vspace{1cm}
\bigskip\hrule

\section{Introduction}
\setcounter{equation}{0}

The absence of observed superpartners for the particles in the Standard Model of particle physics suggests 
that if supersymmetry is a symmetry of Nature, then it is realized in a spontaneously broken form. 
It is somewhat ironic that the second oldest  realization of a supersymmetry algebra in a field-theoretic context 
by Volkov and Akulov \cite{Volkov:1972jx,  Akulov:1974xz} was indeed nonlinear.  

Most of the focus of work on nonlinear realizations of extended supersymmetries has been 
on partially broken global supersymmetry \cite{BW,HLP,APT,FGP,BG,RT,IZ}, 
motivated by the relevance of these constructions to low-energy effective actions for branes in superstring theory.

In this paper we examine fully broken ${\cal N} = 2$ supersymmetry, so that there are two Goldstinos. We use the technique introduced by Ivanov and Kapustnikov 
\cite{Ivanov:1977my,IK1978,IK1982} (see also \cite{Uematsu:1981rj}),  
and further elaborated by Samuel and Wess  \cite{Samuel:1982uh}, to associate a superfield with each Goldstino via a finite supersymmetry transformation with parameters that are the fermionic coordinates of ${\cal N} = 2$ superspace. We find, using standard techniques for construction of nonlinear realizations, that  a special choice of coset parametrization   yields  a version of nonlinearly realized ${\cal N}=2$ supersymmetry for which the associated Goldstino superfields are defined on ${\cal N} = 2$ harmonic superspace, thereby providing a natural mechanism for construction of an ${\cal N} = 2$  Goldstino action.

The plan of the paper is as follows. In section 2, we review the nonlinear realization of ${\cal N} =1$ supersymmetry pioneered by Volkov and Akulov, and the mechanism by which an ${\cal N} =1$ superfield can be associated with the corresponding Goldstino. It is shown that a different choice of coset parametrization from that used by Volkov and Akulov gives rise to an antichiral Goldstino superfield, which allows the construction of a Goldstino action via integration over the antichiral subspace of ${\cal N}= 1$ superspace. The ${\cal N}= 2$ version of this chiral superspace construction is discussed in the appendix. In section 3,  we analyze the case of spontaneously broken ${\cal N} = 2$ supersymmetry, and show that via a particular choice of coset parametrization, it is possible to find Goldstone fields that are naturally adapted to the structure of ${\cal N} = 2$ harmonic superspace.   The operators $D^{++}$ and $D^{--}$ associated with the $SU(2)$ degrees of freedom in harmonic superspace are analyzed in section 4, and it is shown that they have a nonlinear action on the Goldstinos. In section 5, we demonstrate that analytic superfields can be associated with one of the Goldstinos, thus allowing the construction of a Goldstino action by integration over the analytic subspace of ${\cal N} = 2$ superspace. Several reformulations of the Goldstino action are given, and Goldstino-matter couplings 
are introduced.
In section 6,  we briefly discuss composite Goldstino superfields that can be used to construct 
higher-derivative Goldstino actions. Finally, the $\cN=2$ chiral construction is presented in the appendix.

\section{The nonlinear  realizations of $\cN=1$ supersymmetry revisited}
\setcounter{equation}{0}

Nonlinearly realized internal symmetries can be treated systematically using coset constructions \cite{Coleman:1969sm, Callan:1969sn, Isham:1969ci}. Volkov \cite{Volkov} extended the coset construction to include both broken and unbroken spacetime symmetries (see also \cite{Og}), and Volkov and Akulov \cite{Volkov:1972jx,  Akulov:1974xz}  treated the case of broken $\cN$-extended supersymmetries. 

The four-dimensional ${\cal N} = 1$ construction of Volkov and Akulov  \cite{Volkov:1972jx,  Akulov:1974xz} is associated with the supersymmetry algebra
\be
\{ Q_{\a}, \bar{Q}_{\ad} \} = 2 \, P_{\a \ad}~.
\ee
The coset construction of nonlinearly realized supersymmetry on the Goldstone field $\l_{\a}(x)$ is based on the group element
\be
g\big(x, \l (x), \bar{\l} (x)\big) = {\rm e}^{\ri( - x^a P_a + \l^{\a}(x) Q_{\a} + \bar{\l}_{\ad}(x) \bar{Q}^{\ad})}~.
\ee
Supersymmetry transformations are generated by  left action by the group element
\be
g (\e, \bar{\e}) =  {\rm e}^{\ri(\e^{\a} Q_{\a} + \bar{\e}_{\ad} \bar{Q}^{\ad})}~.
\label{ge}
\ee
In infinitesimal form, the supersymmetry transformation law is
\be
\delta \l_{\a} = \e_{\a} - \ri \,v^{\b \bd} \partial_{\b \bd} \l_{\a} ~,\qquad \delta \bar{\l}_{\ad} 
= \bar{\e}_{\ad} - \ri \,  v^{\b \bd} \partial_{\b \bd} \bar{\l}_{\ad}
\ee
with $ v^{\b \bd} = \l^{\b} \bar{\e}^{\bd} - \e^{\b} \bar{\l}^{\bd} .$

An alternative nonlinear realization, first introduced by Zumino  \cite{Zumino:ChiralNLSusy} and further
developed by Samuel and Wess \cite{Samuel:1982uh}, involves a Goldstino $\xi_{\a}$ 
which mixes only with itself under supersymmetry transformations:
\be
\delta \xi_{\a} = \e_{\a} - 2 \ri \, \xi^{\b} \bar{\e}^{\bd} \partial_{\b \bd}  \xi_{\a}.
\label{2.5}
\ee
This nonlinear realization is related to the coset parametrization
\be
g\big(x, \xi (x), \bar{\chi} (x)\big) 
= {\rm e}^{\ri( - x^a P_a + \xi^{\a}(x) Q_{\a})} \, {\rm e}^{\ri  \bar{\psi}_{\ad}(x) \bar{Q}^{\ad}}~.
\ee
Left action by the group element (\ref{ge}) generates the supersymmetry transformation (\ref{2.5}), as well as the transformation
\be
\delta \bar{\psi}_{\ad} = \bar{\e}_{\ad} -  2 \ri \, \xi^{\b} \bar{\e}^{\bd} \partial_{\b \bd}  \bar{\psi}_{\ad} .
\label{2.7}
\ee
The Goldstinos $ \xi_{\a}, \bar{\psi}_{\ad}$ are related to the Volkov-Akulov Goldstinos $\l_{\a}, \bar{\l}_{\ad}$ by 
\be
\xi_{\a} (x) = \l_{\a} (y)~, \qquad \bar{\psi}_{\ad}(x) = \bar{\l}_{\ad} (y)~,
\label{threeGoldstinos}
\ee
where $y^a = x^a - \ri  \l (y) \s^a \bar{\l}(y).$
Conversely,
\be
\l_{\a}(x) = \xi(z)~, \qquad \bar{\l}_{\ad} (x) = \bar{\psi}_{\ad}(z)~,
\label{threeGoldstinos2}
\ee
where $z^a = x^a + \ri \xi(z) \s^a \bar{\psi} (z).$

In the work of Samuel and Wess, the field $\bar{\psi}_{\ad}$ was not exploited. 
Instead, they used $\bar{\xi}_{\ad},$ defined as the Hermitian conjugate of $\xi_{\a},$ 
with the supersymmetry transformation
\be
\delta \bar{\xi}_{\ad} = \bar{\e}_{\ad} + 2 \ri \, \e^{\b} \bar{\xi}^{\bd} \partial_{\b \bd}  \bar{\xi}_{\ad}~.
\ee

However, there is a benefit to be derived from working with the fields $(\xi_{\a}, 
\bar{\psi}_{\ad})$ rather than $(\xi_{\a}, \bar{\xi}_{\ad}).$ 
As first demonstrated by Ivanov and Kapustnikov \cite{Ivanov:1977my,IK1978,IK1982} 
(see also \cite{Samuel:1982uh}), it is possible to associate a superfield with a Goldstino 
via a finite supersymmetry transformation for which the parameter is the corresponding 
fermionic superspace coordinate.
When the Goldstinos  $(\xi_{\a}, 
\bar{\psi}_{\ad})$ are promoted to superfields, the superfield corresponding to  $\bar{\psi}_{\ad}$ is antichiral. This allows the formulation of a Goldstino action by integration over the antichiral subspace of ${\cal N} = 1$ superspace.

Explicitly, the supersymmetry transformations (\ref{2.5}) and (\ref{2.7}) imply
\begin{subequations}
\bea
\ri Q_{\a} \, \xi_{\b} &=&  -  \e_{\a \b }             \\
\ri \bar{Q}_{\ad} \, \xi_{\b} &=& - 2 \ri \, \xi^{\a} \partial_{\a \ad}  \xi_{\b}            
\eea
\end{subequations}
and
\begin{subequations}
\bea
\ri Q_{\a} \, \bar{\psi}_{\bd} &=&   0           \\
\ri \bar{Q}_{\ad} \,  \bar{\psi}_{\bd} &=&   \bar{\e}_{ \ad \bd } - 2 \ri \, \xi^{\a} \partial_{\a \ad} \bar{\psi}_{\bd}~.
\eea
\end{subequations}
The superfields associated with the Goldstinos $\xi_{\a}, \bar{\psi}_{\ad}$ are
\bea
\Xi_{\a} (x, \q, \bar{\q}) = {\rm e}^{\ri X} \xi_{\a} (x), \quad \bar{\Psi}_{\ad} (x, \q, \bar{\q}) 
= {\rm e}^{\ri X} \bar{\psi}_{\ad} (x)~, 
\label{X_L}
\eea
where $X = \q^{\a} Q_{\a} + \bar{\q}_{\ad} \bar{Q}^{\ad}.$
Using 
\be 
D_{\a} {\rm e}^{\ri X} = {\rm e}^{\ri X} \ri Q_{\a}, \quad \bar{D}_{\ad} {\rm e}^{\ri X} 
= {\rm e}^{\ri X} \ri {\bar Q}_{\ad}~,
\ee
it follows that 
\begin{subequations}
\bea
D_{\a} \Xi_{\b} &=&- \e_{\a \b } \\
\bar{D}_{\ad} \Xi_{\b} &=&  - 2 \ri \Xi^{\a} \partial_{\a \ad} \Xi_{\b} 
\eea
\end{subequations}
and
\begin{subequations}
\bea
D_{\a} \bar{\Psi}_{\bd} &=& 0 \label{antichiral} \\
\bar{D}_{\ad} \bar{\Psi}_{\bd} &=&   \e_{ \ad \bd } - 2 \ri \Xi^{\a} \partial_{\a \ad} \bar{\Psi}_{\bd}~ .
\eea
\end{subequations}
For completeness, we also introduce, following Wess and Bagger \cite{Wess,WB},    
the spinor superfield associated with the 
Volkov-Akulov Goldstino $\l_\a$ 
 \bea
\L_{\a} (x, \q, \bar{\q}) = {\rm e}^{\ri X} \l_{\a} (x)~.
\eea
It obeys the constraints \cite{Wess,WB}
 \bea
  \label{AV_Constraints}
	D_\a \L_\b =  - \ve_{ \a \b} + \ri  \bar\L_\ad \pa_\a^\ad\L_\b\,~,
	\qquad 
	{\bar D}_\ad \L_\b = -\ri  \L^\a \pa_{\a \ad} \L_\b ~.
\eea

It can be shown using (\ref{threeGoldstinos2}) that the Goldstino superfields $\X_\a$ and $ \J_\a$ are related to each other as follows:
\bea
\X_\a (z) 
= \J_\b (z) B^\b{}_\a (z) ~, \qquad z^A:= (x^a,\q^\a, \bar \q_\ad ) ~,
\label{2.199}
\eea
for some nonsingular $2\times 2$ matrix $B$ 
such that 
$$
B ={\mathbbm 1} + \mbox{nonlinear Goldstino-dependent terms}~.
$$
 
Eq. (\ref{antichiral}) means 
that the superfield $\bar{\Psi}_\ad$ is indeed antichiral. This allows the construction of an action
\bea
S_\J &=& -\hf \int \rd^4 x \, \rd^2  {\theta} \,{\Psi}^{\a} {\Psi}_{\a} 
-\hf \int \rd^4 x \, \rd^2  \bar{\theta} \,\bar{\Psi}_{\ad} \bar{\Psi}^{\ad} 
\label{2.16} 
\eea
which can be used to describe the Goldstino's dynamics, instead of the action proposed by Samuel
and Wess \cite{Samuel:1982uh}
\bea
S_{\rm SW}= - \int \rd^4 x \, \rd^2 \q \,\rd^2  \bar{\theta}  \, \X^\a \X_\a \, \bar \X_\ad \bar \X^\ad 
\eea
or the superfield version \cite{WB} of the Volkov-Akulov action
\bea
S_{\rm VA}= -\hf \int \rd^4 x \, \rd^2 \q \,\rd^2  \bar{\theta}  \, \L^\a \L_\a  \, \bar \L_\ad \bar \L^\ad ~.
\eea

The component form of the action (\ref{2.16})  is 
\bea
S_\J&=&-  \int \rd^4 x \,\Big(\hf +  \ri \, \xi^{\a} \partial_{\a \ad} \bar{\psi}^{\ad} 
-  \xi^{\a} (\partial_{\a \ad} \xi_{\b}) \partial^{\ad \b} \bar{\psi}^2 
- \frac14 \xi^2 \partial^{\ad \a} \partial_{\a \ad} \bar{\psi}^2 +{\rm c.c.}\Big)~.~~~
\label{2.21}
\eea
The striking feature of this action is that it is at most quartic in the Goldstino fields, 
unlike the other Goldstino models which in general contain 
 terms to eighth order in the fermionic field\footnote{The Volkov-Akulov action
does not  contain any terms of eighth  order in $\l$ and $\bar{\l}$ \cite{KMcC}.}
 (see \cite{KT} for a detailed 
 analysis of the known Goldstino models and their relationships). 
This unusual feature is deceptive, however,
since the action is given in terms of two different sets of fields, 
$(\x_\a , \bar \x^\ad )$ and $(\j_\a , \bar \j^\ad )$ related  to each other as in via (\ref{threeGoldstinos}, \ref{threeGoldstinos2}).
Sixth- and eighth-order terms will inevitably appear once the action is expressed via one set of fields or the other.
If we express the action entirely in terms of the fields $(\x_\a , \bar \x^\ad )$, we must end up with 
(the component version of) the Samuel-Wess Goldstino action. On the other hand,
expressing (\ref{2.21}) entirely in terms of  the fields $(\j_\a , \bar \j^\ad )$, 
it turns out that we end up with Ro\v{c}ek's  Goldstino superfield \cite{Rocek}. 
This claim can be justified as follows.

We can associate with the Goldstino superfields two composite {\it nilpotent} objects
which contain all the information about the original superfields, specifically: 
\bea
\F &:=& \J^\a \J_\a~;  \label{2.24}\\
\S &:=&  {\bar \X}_\ad {\bar \X}^\ad ~.
\eea
The superfield $\S$ is equivalent to the complex linear Goldstino superfield introduced in  \cite{Kuzenko:2011ti}. 
Its properties are
\bea
 -\frac{1}{4}  {\bar D}^2 \S=1~, \qquad \S^2=0~, \qquad -\frac{1}{4} \S{\bar D}^2D_\a\S &=  D_\a\S~.
\label{2.26}
 \eea
The superfield $\F$ proves to be  equivalent to  Ro\v{c}ek's chiral Goldstino superfield \cite{Rocek} as 
its properties are 
\bea
{\bar D}_\ad \F =0 ~, \qquad \F =0 ~, \qquad
-\frac{1}{4}\F{\bar D}^2 \overline{ \F} = \F~.
\label{2.27}
\eea
Eq. (\ref{2.199}) has to be used to derive the last constraint in (\ref{2.27}).
This confirms that (\ref{2.16}) is equivalent to Ro\v{c}ek's Goldstino action
\bea
S_{\rm R} &=& - \int \rd^4 x \, \rd^2\q\, \F=  
- \int \rd^4 x \, \rd^2 \q \,\rd^2  \bar{\theta}  \, \bar \F \F~.
\eea
Our realization (\ref{2.16}) shows that $S_\J$ may be interpreted as a square root of 
$S_{\rm R}$.

Using eqs. (\ref{2.26}) and (\ref{2.27}), it is not difficult to show that the Goldstino action (\ref{2.16})
can be expressed only in terms of the Goldstone superfields (\ref{X_L}) as 
\bea
S_{\X \bar \J}= - \int \rd^4 x \, \rd^2 \q \,\rd^2  \bar{\theta}  \, \X^\a \X_\a \, \bar \J_\ad \bar \J^\ad ~.
\label{2.29}
\eea
This action proves to be  real, as a consequence of the third constraint in (\ref{2.27}).

\section{Analytic  realization of spontaneously broken  ${\cal N} = 2$ supersymmetry}
\setcounter{equation}{0}

In the case of $\cN=2$ supersymmetry, one can also introduce a  nonlinear realization that gives rise to Goldstino superfields that are antichiral, as explicitly described in the Appendix. 
However, for reasons discussed above in the $\cN=1$ case, such a realization appears 
to be less useful than the $\cN=2$  analogue of  the construction by Samuel and Wess
 \cite{Samuel:1982uh}.
This motivates seeking a new approach
to the description of spontaneously broken ${\cal N} = 2$ supersymmetry.

Harmonic superspace \cite{Galperin:1984av}
(see \cite{Galperin:2001uw} for a review)
extends conventional ${\cal N} = 2$ superspace by the two-sphere $S^2 = {\rm SU(2)/U(1)}$ parametrized by the group elements
$$
(u_i{}^{-}, u_i{}^{+} ) \in {\rm SU}(2)~, \qquad u_{i}^+ := \e_{ij}u^{+j}~, \qquad 
\overline{u^{+i} } = u_{i}^-~, \qquad u^{+i}u_i^- = 1~.
$$
The ${\cal N} = 2$ supersymmetry algebra, 
\be
\{ Q_{\a}^i, \bar{Q}_{\ad j} \} = 2 \, \d^i_j P_{\a \ad}~,
\label{N=2}
\ee
can be re-cast in the form
\be
\{ Q_{\a}^{\pm} , \bar{Q}_{\ad}^{\mp} \} = \pm \, 2 \, P_{\a \ad}~,
\ee
where  
\be 
Q_{\a}^{\pm} = Q_{\a}^{i} u_i^{\pm}, \quad \bar{Q}_{\ad}^{\pm} =\bar{Q}_{\ad}^{i}  u_i^{\pm} ~.
\ee

To examine the complete breaking of ${\cal N} = 2$ supersymmetry in a harmonic superspace context, 
we choose a very particular coset parametrization
\be
g(x, \l (x), \bar{\l} (x)) = {\rm e}^{\ri(-x^aP_a + \l^{\a -}(x) \,Q_{\a}^+ - \bar{\l}_{\ad}^-(x) \,\bar{Q}^{\ad +})} \, 
{\rm e}^{\ri(- \l^{\a +}(x)\, Q_{\a}^- + \bar{\l}_{\ad}^+(x)\, \bar{Q}^{\ad -})}~.
\label{coset}
\ee
The fields in (\ref{coset}) are related to the Volkov-Akulov Goldstinos, eq. (\ref{A.1}), as follows:
\bea
\l^{\a}_i (x ' ) &=& \l^{\a -}(x) u_i^+ - \l^{\a +} (x) u_i^- ~, \qquad
\bar{\l}^{\ad}_i (x ' ) = \bar{\l}^{\ad -}(x) u_i^+ - \bar{\l}^{\ad +} (x) u_i^- ~,
\eea
where
\bea
x^{ \prime a} &=& x^a - \ri \l^-(x) \s^a \bar{\l}^+(x) - \ri \l^+(x) \s^a \bar{\l}^- (x)~.
\eea
Left action by the group element 
$$ 
g(\e, \bar{\e}) =
{\rm e}^{\ri(\e^{\a -} \,Q_{\a}^+ - \bar{\e}_{\ad}^- \,\bar{Q}^{\ad +}- \e^{\a +}\, Q_{\a}^- + \bar{\e}_{\ad}^+\, \bar{Q}^{\ad -})}
$$
gives rise to the infinitesimal supersymmetry transformations
\begin{subequations}
\bea
\d \l_{\a}^+ &=& \e_{\a}^+  - 2 \ri \, \e^{\b +} \bar{\l}^{\bd -} \partial_{\b \bd} \l_{\a}^+  + 2\ri \, \bar{\e}^{\bd +} \l^{\b -} \partial_{\b \bd} \l_{\a}^+ \\
\d \bar{\l}_{\ad}^+ &=& \bar{\e}_{\ad}^+ - 2 \ri \, \e^{\b +} \bar{\l}^{\bd -} \partial_{\b \bd}  \bar{\l}_{\ad}^+  
+ 2\ri \, \bar{\e}^{\bd +} \l^{\b -}  \partial_{\b \bd}  \bar{\l}_{\ad}^+ \\
\d \l_{\a}^- &=& \e_{\a}^- - 2 \ri \, \e^{\b +} \bar{\l}^{\bd -} \partial_{\b \bd} \l_{\a}^-  + 2\ri \,\bar{\e}^{\bd +}  \l^{\b -}  \partial_{\b \bd} \l_{\a}^-  \\
\d \bar{\l}_{\ad}^- &=& \bar{\e}_{\ad}^- - 2 \ri \, \e^{\b +} \bar{\l}^{\bd -} \partial_{\b \bd}  \bar{\l}_{\ad}^-  
+ 2\ri \, \bar{\e}^{\bd +}  \l^{\b -} \partial_{\b \bd}  \bar{\l}_{\ad}^- ~. 
\eea
\end{subequations}
The significance of 
the coset parametrization chosen 
is that $\d \l_{\a}^+ $ and $\d \bar{\l}_{\ad}^+$ have no dependence on $\e^{\a -}$ and $\bar{\e}_{\ad}^{\, -},$ meaning that $\l_{\a}^+$ and $\bar{\l}_{\ad}^+$ are annihilated by $Q_{\a}^+$ and $\bar{Q}^{\ad +}$: 
\begin{subequations}
\bea
Q_{\b}^+ \, \l_{\a}^+ &=& 0 \\
\bar{Q}_{\bd}^+ \, \l_{\a}^+ &=& 0 \\
Q_{\b}^- \, \l_{\a}^+ &=& \ri \,\e_{\a \b} + 2  \bar{\l}^{\bd -} \partial_{\b \bd} \l_{\a}^+ \\ 
\bar{Q}_{\bd}^- \, \l_{\a}^+ &=&  - 2 \l^{\b -} \partial_{\b \bd} \l_{\a}^+ 
\eea
\end{subequations}
and
\begin{subequations}
\bea
Q_{\b}^+ \bar{\l}_{\ad}^+ &=& 0 \\
\bar{Q}_{\bd}^+\bar{\l}_{\ad}^+ &=& 0 \\
Q_{\b}^- \bar{\l}_{\ad}^+ &=&  2 \bar{\l}^{\bd -} \partial_{\b \bd} \bar{\l}_{\ad}^+\\ 
\bar{Q}_{\bd}^- \bar{\l}_{\ad}^+ &=& \ri \,\e_{\ad \bd} - 2 \l^{\b -} \partial_{\b \bd} \bar{\l}_{\ad}^+~ .
\eea
\end{subequations}
When we come to construct superfields from $\l_{\a}^+$ and $\bar{\l}_{\ad}^+,$ this will ensure that the corresponding superfields are analytic.

The supersymmetry transformations of $\l_{\a}^-$ and $\bar{\l}_{\ad}^-$ are given by
\begin{subequations}
\bea
Q_{\b}^+ \l_{\a}^- &=& -\ri \,\e_{\a \b} \\
\bar{Q}_{\bd}^+ \l_{\a}^- &=& 0 \\
Q_{\b}^- \l_{\a}^- &=&   2 \bar{\l}^{\bd -} \partial_{\b \bd} \l_{\a}^- \\ 
\bar{Q}_{\bd}^- \l_{\a}^- &=&  - 2 \l^{\b -} \partial_{\b \bd} \l_{\a}^- 
\eea
\end{subequations}
and
\begin{subequations}
\bea
Q_{\b}^+ \bar{\l}_{\ad}^- &=& 0 \\
\bar{Q}_{\bd}^+\bar{\l}_{\ad}^- &=& -\ri \,\e_{\ad \bd} \\
Q_{\b}^- \bar{\l}_{\ad}^- &=&  2\ri \, \bar{\l}^{\bd -} \partial_{\b \bd} \bar{\l}_{\ad}^-\\ 
\bar{Q}_{\bd}^- \bar{\l}_{\ad}^- &=&  - 2\ri \, \l^{\b -} \partial_{\b \bd} \bar{\l}_{\ad}^- ~.
\eea
\end{subequations}

It should be pointed out that the fields $\l^+_\a$ and ${\bar \l}^+_\ad $ are conjugate 
with respect to the analyticity-preserving conjugation \cite{Galperin:1984av}: 
\begin{subequations}
\bea
\widetilde{\l^+_\a} = {\bar \l}^+_\ad~, \qquad \widetilde{\bar \l^+_\ad} = -{ \l}^+_\a~.
\eea
Similar relations hold for the fields $\l^-_\a$ and ${\bar \l}^-_\ad $, that is  
\bea
\widetilde{\l^-_\a} = {\bar \l}^-_\ad~, \qquad \widetilde{\bar \l^-_\ad} = -{ \l}^-_\a~.
\eea
\end{subequations}
This is the key feature distinguishing the nonlinear realization in this section from the antichiral one detailed in the Appendix. In the latter, there is no simple conjugation relation between the Goldstinos associated with dotted and undotted supersymmetry generators (unlike the Volkov-Akulov realization (\ref{A.1})). 
In the parametrization  (\ref{coset})  adapted to harmonic superspace, 
Goldstinos are related by analyticity-preserving conjugation.

\section{Including the harmonic operators $D^{++}$ and $D^{--}$}
\setcounter{equation}{0}

In  harmonic superspace, a key role is played by the SU(2) left-invariant vector fields
\be
D^{++} = u^{+i} \frac{\partial}{\partial u^{-i}}~, \quad 
D^{--} = u^{-i} \frac{\partial}{\partial u^{+i}}~, 
\quad D^0 = u^{+i} \frac{\partial}{\partial u^{+i}} - u^{-i} \frac{\partial}{\partial u^{-i}}~.
\ee
Here $D^0$ is the operator associated with harmonic U(1) charge. Given a scalar function $\vf^{(n)}(u^+, u^-)$
on SU(2) of U(1) charge $n$,
\bea
\vf^{(n)}( {\rm e}^{ {\rm i}\a}\, u^+, {\rm e}^{ -{\rm i}\a}\,u^-)
= {\rm e}^{ {\rm i}n\a} \,\vf^{(n)}(u^+, u^-)~, \qquad \a \in {\mathbb R}~,
\eea
we have $D^0 \vf^{(n)} = n \vf^{(n)}$. 
Of particular importance for in our considerations is the operator $D^{++}$ which 
acts on the supersymmetry generators as
\be
[ D^{++}, Q_{\a}^+ ] = 0~, \quad   [ D^{++}, Q_{\a}^- ] = Q_{\a}^+~,  \quad [ D^{++}, \bar{Q}_{\ad}^+ ] = 0~, 
\quad [ D^{++}, \bar{Q}_{\ad}^- ] = \bar{Q}_{\ad}^+~. 
\ee 

At first glance, one might expect 
 \be
 D^{++}\l_{\a}^+  = 0~, \quad D^{++} \l_{\a}^-  = \l_{\a}^+~,  \quad D^{++}\bar{\l}_{\ad}^+  = 0~,  
 \quad D^{++} \bar{\l}_{\ad}^-  = \bar{\l}_{\ad}^+~.
\ee
However, these are inconsistent with the nonlinear action of the supersymmetry generators on the Goldstinos. We can see this with a simple example. Assuming $[ D^{++}, Q_{\b}^- ] = Q_{\b}^+ $ and the nonlinear supersymmetry transformations of the $\l's,$
\bea
0 &=& Q_{\b}^+ \l_{\a}^+ \nonumber \\
&=& [ D^{++}, Q_{\b}^- ] \l_{\a}^+  \nonumber  \\
&=& D^{++} (-\ri \e_{\a \b} + 2 \bar{\l}^{\bd -} \partial_{\b \bd} \l_{\a}^+) - Q_{\b}^- (D^{++}  \l_{\a}^+) \nonumber  \\
&=& 2 (D^{++}  \bar{\l}^{\bd -}) \partial_{\b \bd} \l_{\a}^+ + 2 \bar{\l}^{\bd -} \partial_{\b \bd}( D^{++} \l_{\a}^+) -  Q_{\b}^- (D^{++}  \l_{\a}^+)~, \nonumber 
\eea
which yields a contradiction if we set $D^{++}\l_{\a}^+  = 0.$

In order to determine how the operator 
$D^{++}$ acts on the Goldstinos, note that the action of the supersymmetry generators  
on the Goldstinos can  be represented {\it schematically} in the form:
\begin{subequations}
\bea
Q_{\b}^+  &=& -\ri \,\frac{\partial}{\partial \l^{\b -}} \\
\bar{Q}_{\bd}^+  &=&  -\ri \, \frac{\partial}{\partial \bar{ \l}^{\bd -}}  \\
Q_{\b}^- &=&   \ri \, \frac{\partial}{\partial \l^{\b +}} + 2 \bar{\l}^{\bd -} \partial_{\b \bd}  \\ 
\bar{Q}_{\bd}^-  &=&  \ri \, \frac{\partial}{\partial \bar{ \l}^{\bd +}}  - 2 \l^{\b -} \partial_{\b \bd} ~ .
\eea
\end{subequations}
This is similar to the superspace representation of the supersymmetry generators in the analytic basis, 
and the corresponding central basis version is 
\bea
\widehat{Q} = {\rm e}^Z Q {\rm e}^{-Z}~, \qquad
Z := - \ri \l^{\b -} \bar{\l}^{\bd +} \partial_{\b \bd} - \ri  \l^{\b +} \bar{\l}^{\bd -} \partial_{\b \bd} ~.
\eea
One finds
\begin{subequations}
\bea
\widehat{Q}^+  &=& -\ri \,\frac{\partial}{\partial \l^{\b -}} +  \bar{\l}^{\bd +} \partial_{\b \bd} \ \\
\widehat{Q}_{\bd}^+  &=&  -\ri \, \frac{\partial}{\partial \bar{ \l}^{\bd -}} - \l^{\b +} \partial_{\b \bd}   
\\
\widehat{Q}_{\b}^- &=&   \ri \, \frac{\partial}{\partial \l^{\b +}} +  \bar{\l}^{\bd -} \partial_{\b \bd}  \\ 
\widehat{Q}_{\bd}^-  &=&  \ri \,\frac{\partial}{\partial \bar{ \l}^{\bd +}}  - \l^{\b -} \partial_{\b \bd} ~ .
\eea
\end{subequations}
In this central basis, $\widehat{D}^{++} = u^{+i} \frac{\partial}{\partial u^{-i}} $ satisfies 
$$ [ \widehat{D}^{++}, \widehat{Q}^+ ] = 0, \qquad  
[ \widehat{D}^{++}, \widehat{Q}^- ] = \widehat{Q}^+ .$$
It follows that in the analytic basis,
\bea
 D^{++} &=& e^{-Z}  \widehat{D}^{++} e^Z 
 =  \widehat{D}^{++} - 2 \ri \l^{\b +} \bar{\l}^{\bd +} \partial_{\b \bd}~.
 \eea 
 Thus  $[ D^{++}, Q_{\a}^+ ] = 0, [ D^{++}, Q_{\a}^- ] = Q_{\a}^+$ etc, and 
 \begin{subequations}\label{3.8}
\bea
D^{++} \l_{\a}^+  &=& - 2\ri \,\l^{\b+}\bar{\l}^{\bd +} \partial_{\b \bd} \l_{\a}^+  \\
D^{++} \l_{\a}^-  &=& \l_{\a}^+ - 2\ri \, \l^{\b+}\bar{\l}^{\bd +} \partial_{\b \bd} \l_{\a}^-  \label{4.8b}  \\
D^{++} \bar{ \l}_{\ad}^+  &=& - 2\ri \, \bar{ \l}^{\b+}\bar{\l}^{\bd +} \partial_{\b \bd}  \bar{ \l}_{\ad}^+   \\
D^{++} \bar{ \l}_{\ad}^-  &=&  \bar{ \l}_{\ad}^+ - 2\ri \, \l^{\b+}\bar{\l}^{\bd +} \partial_{\b \bd}  \bar{ \l}_{\ad}^-~ .
\label{4.8c}
\eea
\end{subequations}
In particular, this shows that the Goldstinos $\l_{\a}^{\pm}$ and $\bar{\l}_{\ad}^{\pm}$ 
in the coset parametrization (\ref{coset}) are not simply of the form $\l_{\a}^{i} u_{i}^{\pm}$ 
and $\bar{\l}_{\ad}^{i} u_i^{\pm},$ with $\l_{\a}^{i} $ and $\bar{\l}_{\ad}^{i} $ conventional 
${\cal N} = 2$ Goldstinos independent of the harmonic coordinates.

The equations (\ref{4.8b}) and (\ref{4.8c}) are very important, 
since they explicitly relate the Goldstino fields $(\l^-_\a, \bar \l^-_\ad )$ to $(\l^+_\a, \bar \l^+_\ad )$.

In a similar manner, one can prove
\begin{subequations}
\bea
D^{--} \l_{\a}^+  &=& \l_{\a}^- - 2\ri \, \l^{\b-}\bar{\l}^{\bd -} \partial_{\b \bd} \l_{\a}^+   \\
D^{--} \l_{\a}^-  &=&  - 2\ri \, \l^{\b-}\bar{\l}^{\bd -} \partial_{\b \bd} \l_{\a}^-   \\
D^{--} \bar{ \l}_{\ad}^+  &=& \bar{ \l}_{\ad}^- - 2\ri \, \bar{ \l}^{\b-}\bar{\l}^{\bd -} \partial_{\b \bd}  \bar{ \l}_{\ad}^+   \\
D^{--} \bar{ \l}_{\ad}^-  &=&   - 2\ri \, \l^{\b-}\bar{\l}^{\bd -} \partial_{\b \bd}  \bar{ \l}_{\ad}^- ~.
\eea
\end{subequations}

It follows from (\ref{3.8}) that 
\bea
(D^{++})^4 \l_\a^+ =0~, 
\qquad (D^{++})^4 {\bar \l}^+_\ad=0~, 
\eea
and therefore $\l^+_\a$ and $\bar \l^+_\ad$ have  simple harmonic expansions 
in powers of $u^+_i $ and $u^-_i$, in particular
\bea
\l^+_\a (u^+,u^-)& =& 
\l_\a^i u^+_i +\sum_{n=1}^{3} 
\l_\a^{( i_1 \dots i_{n+1} j_1\dots j_n )} u^+_{i_1} \dots u^+_{i_{n+1}} u^-_{j_1} \dots u^-_{j_n}~.
\label{4.11}
\eea
Here $\l^i_\a$ is the $\cN=2$ Volkov-Akulov Goldstino, eq. (\ref{A.1}).
The other fields in (\ref{4.11}) are nonlinear functions of $\l^\a_i$ and its conjugate.

\section{Construction of Goldstino superfields}
\setcounter{equation}{0}

Similarly to  the $\cN=1$ case discussed in section 2, 
to each of  the Goldstinos $\l_{\a}^{\pm}$ and $\bar{\l}_{\ad}^{\pm}$  we can associate an ${\cal N}=2$ superfield depending on  fermionic superspace coordinates $\q_{\a}^{\pm}$ and $\bar{\q}_{\ad}^{\pm}$. In general form,
\bea
 \L(x, \q, \bar{\q}) = {\rm e}^{\ri X
} \, \l (x)~, \qquad
X:= \q^{\a -} Q_{\a}^+ - \bar{\q}_{\ad}^-\bar{Q}^{\ad +} 
- \q^{\a +} Q_{\a}^- + \bar{\q}_{\ad}^+ \bar{Q}^{\ad -}~.
\eea

In the central basis for harmonic superspace\footnote{Our use of harmonic-superspace terminology is 
somewhat unorthodox.},  
the supercovariant derivatives take the form
\bea
D_{\b}^+ &=&   \phantom{-}  \frac{\partial}{\partial \q^{\b -}} + \ri \, \bar{\q}^{\bd +} \partial_{\b \bd} ~,\qquad
\bar{D}_{\bd}^+ =   \phantom{-}  \frac{\partial}{\partial \bar{\q}^{\bd -}} - \ri \, \q^{\b +} \partial_{\b \bd} \nonumber \\
D_{\b}^- &=&    - \frac{\partial}{\partial \q^{\b +}} + \ri  \,\bar{\q}^{\bd -} \partial_{\b \bd} 
~, \qquad
\bar{D}_{\bd}^- =    - \frac{\partial}{\partial \bar{\q}^{\bd +}} - \ri \, \q^{\b -} \partial_{\b \bd}    \nonumber
\eea
and satisfy the algebra
\be
\{ D_{\b}^{\pm}, \bar{D}_{\bd}^{\mp} \} = \mp \, 2 \ri \, \partial_{\b \bd}~.
\ee
The action of the spinor covariant derivatives on 
the Goldstino superfields\footnote{The construction $\L (x, \q, \bar{\q}) = {\rm e}^{\ri X} \l(x)$ with 
$ X = \q^{\a -} Q_{\a}^+ - \bar{\q}_{\ad}^-\bar{Q}^{\ad +} - \q^{\a +} Q_{\a}^- + \bar{\q}_{\ad}^+ \bar{Q}^{\ad -}$ is adapted to the central basis for the supercovariant derivatives. The analytic basis is adapted to
$\L (x, \q, \bar{\q}) = {\rm e}^{\ri X_1} {\rm e}^{\ri X_2} \l(x)$ with
$ X_1 = - \q^{\a +} Q_{\a}^- + \bar{\q}_{\ad}^+ \bar{Q}^{\ad -}$ and
$ X_2 = \q^{\a -} Q_{\a}^+ - \bar{\q}_{\ad}^-\bar{Q}^{\ad +}.$} 
replicates the action of the supersymmetry generators on the Goldstinos via
\bea
D_{\b}^{\pm} \L &=&  {\rm e}^{\ri X}
\, \ri \, Q_{\b}^{\pm} \l  ~, \qquad
\bar{D}_{\bd}^{\pm} \L =  {\rm e}^{\ri X}
\, \ri \,\bar{Q}_{\bd}^{\pm} \l ~.
\eea

In particular we obtain
\begin{subequations}
\bea
D_{\b}^+ \L_{\a}^+ &=& 0 \\
\bar{D}_{\bd}^+ \L_{\a}^+ &=& 0 \\
D_{\b}^- \L_{\a}^+ &=& - \e_{\a \b} + 2\ri \, \bar{\L}^{\bd -} \partial_{\b \bd} \L_{\a}^+ \\ 
\bar{D}_{\bd}^- \L_{\a}^+ &=&  - 2\ri \, \L^{\b -} \partial_{\b \bd} \L_{\a}^+~ ,
\eea
\end{subequations}
meaning that $ \L_{\a}^+ $ is an analytic superfield. Also
\begin{subequations}
\bea
D_{\b}^+ \bar{\L}_{\ad}^+ &=& 0 \\
\bar{D}_{\bd}^+\bar{\L}_{\ad}^+ &=& 0 \\
D_{\b}^- \bar{\L}_{\ad}^+ &=& 2\ri \, \bar{\L}^{\bd -} \partial_{\b \bd} \bar{\L}_{\ad}^+\\ 
\bar{D}_{\bd}^- \bar{\L}_{\ad}^+ &=& - \e_{\ad \bd} - 2\ri \, \L^{\b -} \partial_{\b \bd} \bar{\L}_{\ad}^+ ~,
\eea
\end{subequations}
so $ \bar{\L}_{\ad}^+$  is also an analytic superfield.

The remainder of the $D$-algebra is given by
\begin{subequations}
\bea
D_{\b}^+ \L_{\a}^- &=& \e_{\a \b} \\
\bar{D}_{\bd}^+ \L_{\a}^- &=& 0 \\
D_{\b}^- \L_{\a}^- &=&   2\ri \, \bar{\L}^{\bd -} \partial_{\b \bd} \L_{\a}^- \\ 
\bar{D}_{\bd}^- \L_{\a}^- &=&  - 2\ri \, \L^{\b -} \partial_{\b \bd} \L_{\a}^- 
\eea
\end{subequations}
and
\begin{subequations}
\bea
D_{\b}^+ \bar{\L}_{\ad}^- &=& 0 \\
\bar{D}_{\bd}^+\bar{\L}_{\ad}^- &=& \e_{\ad \bd} \\
D_{\b}^- \bar{\L}_{\ad}^- &=&  2\ri \, \bar{\L}^{\bd -} \partial_{\b \bd} \bar{\L}_{\ad}^-\\ 
\bar{D}_{\bd}^- \bar{\L}_{\ad}^- &=&  - 2\ri \, \L^{\b -} \partial_{\b \bd} \bar{\L}_{\ad}^- ~.
\eea
\end{subequations}

To find the action of the $SU(2)$ generator $D^{++}$ on the Goldstino superfields, represented schematically in the form $ \L (x, \q, \bar{\q}) = {\rm e}^{\ri X}  \l(x)$ with $ X = \q^{\a -} Q_{\a}^+ - \bar{\q}_{\ad}^-\bar{Q}^{\ad +} - \q^{\a +} Q_{\a}^- + \bar{\q}_{\ad}^+ \bar{Q}^{\ad -},$ we note $D^{++} X = 0,$ so 
$$ D^{++} \L (x, \q, \bar{\q}) = {\rm e}^{\ri X} D^{++} \l(x).$$
Thus the transformation properties of the superfield $\L (x, \q, \bar{\q})$ are determined by those of the corresponding field $\l(x),$
\begin{subequations}\label{4.7}
\bea
D^{++}  \L_{\a}^+ &=&  - 2\ri \, \L^{\b +} \bar{\L}^{\bd +} \partial_{\b \bd} \L_{\a}^+  \label{4.7a} \\
D^{++}  \L_{\a}^- &=& \L_{\a}^+ -  2\ri \, \L^{\b +}\bar{\L}^{\bd +}\partial_{\b \bd} \L_{\a}^-  \\
D^{++} \bar{\L}_{\ad}^+ &=&  - 2\ri \, \L^{\b +} \bar{\L}^{\bd +} \partial_{\b \bd} \bar{\L}_{\ad}^+ 
\label{4.7c}  \\
D^{++}  \bar{ \L}_{\ad}^- &=& \bar{ \L}_{\ad}^+ - 2\ri \, \L^{\b +} \bar{\L}^{\bd +} \partial_{\b \bd} \bar{\L}_{\ad}^-~.
\eea
\end{subequations}
Similarly, one can prove
\begin{subequations}\label{4.8}
\bea
D^{--}  \L_{\a}^+ &=& \L_{\a}^-  - 2\ri \, \L^{\b -} \bar{\L}^{\bd -} \partial_{\b \bd} \L_{\a}^+  \\
D^{--}  \L_{\a}^- &=&  -  2\ri \, \L^{\b -} \bar{\L}^{\bd -}  \partial_{\b \bd} \L_{\a}^-  \\
D^{--} \bar{\L}_{\ad}^+ &=& \bar{ \L}_{\ad}^-  - 2\ri \, \L^{\b -} \bar{\L}^{\bd -}  \partial_{\b \bd} \bar{\L}_{\ad}^+  \\
D^{--}  \bar{ \L}_{\ad}^- &=& - 2\ri \, \L^{\b -} \bar{\L}^{\bd -} \partial_{\b \bd} \bar{\L}_{\ad}^-~.
\eea
\end{subequations}

Since $\L_{\a}^+$ and $\bar{\L}_{\ad}^+$ are analytic superfields, 
it is possible to  consistently define an action
\bea
S =\int {\rm d}u  \int {\rm d} \z^{(-4)} \, L^{(+4)} ~, \qquad 
L^{(+4)}= -\hf  \L^{\a+} \L_{\a}^+
\bar{\L}_{\ad}^+   \bar{\L}^{\ad + }~,
\label{action}
\eea
where the integration is over the analytic subspace of harmonic superspace, 
\bea
  {\rm d} \z^{(-4)}: = 
 {\rm d}^4 x 
  \,  (D^{-})^4~, \qquad 
 (D^-)^{4}
 :=
 \frac{1}{16} 
 (\bar{D}^-)^2 (D^-)^2 ~.
 \eea
 In (\ref{action}), 
 $ \int {\rm d}u $ denotes the integration over the group manifold SU(2) defined as in  
  \cite{Galperin:2001uw} 
\be
\int {\rm d}u \; 1 = 1 \qquad \int {\rm d}u \, u^+_{(i_1}
\ldots u^+_{i_n} u^-_{j_1}
\ldots u^-_{j_m)} = 0 \qquad n+m > 0\;.
\label{hint}
\ee  
Direct calculations give 
\bea
{} & & \frac{1}{16} \int {\rm d}u\, (\bar{D}^-)^2 (D^-)^2 \, (\bar{\L}^+)^2 (\L^+)^2| 
= 1 + \ri \, \l^{\a }_i \partial_{\a \ad} \bar{\l}^{\ad i} + \ri \, \bar{\l}^{\ad  i} \partial_{\a \ad} \l^{\a }_i 
+O(\l^4)~,~~~
 \eea
and thus the above action generates the correct kinetic term for the Goldstinos.

It is important to point out that  the equations (\ref{4.7a}) and (\ref{4.7c}) imply
the following nontrivial property
\bea
D^{++} L^{(+4)}=0 \quad \Longleftrightarrow \quad L^{(+4)} (u) = L^{ijkl} u^+_i u^+_j u^+_k u^+_l~,
\eea
so that the Lagrangian is in fact independent of $u_i^-$.
Since $L^{(+4)}$ is analytic,  the superfield  $L^{ijkl}$ obeys the analyticity constraints
\bea
D_\a^{(m} L^{ijkl)} = {\bar D}_\ad^{(m} L^{ijkl)} =0~.
\label{5.12}
\eea
Following the terminology \cite{G-RRWLvU} of $\cN=2$ projective superspace 
\cite{LR-projective} (see \cite{K10} for a modern review), 
the Lagrangian $L^{(+4)} $ is an $\cO (4)$ multiplet \footnote{The 
fact that $L^{(+4)}$ is a holomorphic tensor field over ${\mathbb C}P^1$, i.e. independent of $u^-_i$, 
 may be of significance in formulating a curved superspace version of the model under consideration.}.
We can now reformulate the Goldstino action in projective superspace\footnote{The relationship between 
the $\cN=2$ harmonic and projective superspace approaches is spelled out in \cite{K10,K98}.}.
To pass from the harmonic to the projective formulation requires several steps.   First of all, 
one should replace $ u^{+i} \to v^i \in {\mathbb C}^2 \setminus \{0\}$
where  the isotwistor $v^i$  provides homogeneous coordinates for ${\mathbb C}P^1$.
Secondly, one should also replace $u^-_i \to w_i $, where the isotwistor $w_i$ is subject only to the condition 
$(v,w):=v^iw_i \neq 0$, and is otherwise  completely arbitrary. 
Thirdly, the integration over SU(2) in (\ref{action}) should be replaced by a closed contour integral
in ${\mathbb C}P^1$. The projective-superspace realization of the Goldstino action 
is as follows\footnote{The action (\ref{projective-action}) can be obtained from (\ref{action}) 
by applying  the reduction technique developed in \cite{K98}.};
\bea
S= -\frac{c}{ 2\pi}
\oint 
\frac{v_i \rd v^{i}}{ (v, w)^4}\int\rd^4 x\,({D}^-)^4 \frac{ L^{(+4)}(v) }{c^{(+2)}(v)}  ~, 
\qquad  c^{(+2)}(v) := c^{ij} v_i v_j~,
\label{projective-action}
\eea
with $c^{ij}= c^{ji}$ a real non-zero {\it constant} isovector, and $2c^2 =c^{ij} c_{ij}$. 
One can show that this action is 
independent of $w_i$ and $c^{ij}$.

The properties of $L^{(+4)}$ given  above are such that 
the action (\ref{action}) can be brought to the form:
\bea
S=\frac{1}{80} \int  {\rm d}^4 x \,D^{ i} D^j {\bar D}^k {\bar D}^{ l} L_{ijkl}|~.
\label{5.13}
\eea
Thirty years ago, 
Sohnius, Stelle and West \cite{SSW} realized that one can associate 
an $\cN=2$ super-action of the form (\ref{5.13})
with any real symmetric iso-tensor 
superfield $L^{ijkl}$ under the constraints (\ref{5.12}). 
It is interesting that their supersymmetric action principle
can be used to describe spontaneous supersymmetry breaking. 

The Goldstino action, eq. (\ref{action}),  can also be represented 
as an integral over full $\cN=2$ superspace
\bea
S= \int \rd^4x \, \rd^4 \q \, \rd^4 {\bar \q}\, L~, \qquad
L := -\hf ( \L^+)^2 (\bar \L^+ )^2  ( \L^-)^2 (\bar \L^- )^2 ~.
\label{5.17}
\eea
There is no need to include the integration over SU(2), since the Lagrangian $L$ 
is independent of the harmonics, 
\be
D^{++}L=D^{--}L=0~,
\ee
as follows from (\ref{4.7}) and (\ref{4.8}).

We have constructed three different versions of the Goldstino action:
(i) the harmonic superspace realization (\ref{action}); (ii) the projective superspace realization
(\ref{projective-action}); (iii) the full superspace realization (\ref{5.17}).
They allow us to generate three different types of Goldstino-matter couplings. 
Within the harmonic-superspace approach, couplings of the Goldstinos to $\cN=2$ supersymmetric 
matter can be described by actions of the form:
\bea
S_{\text{inter, harmonic}} =\int {\rm d}u  \int {\rm d} \z^{(-4)} \, L^{(+4)} (u^+)\, \cL^{(0)}_{\rm matter} (u^+, u^-)~,
\label{5.20}
\eea
where $ \cL^{(0)}_{\rm matt}$ is a real analytic superfield of U(1) charge zero, 
\bea
D^+_\a  \cL^{(0)}_{\rm matter} = {\bar D}^+_\ad  \cL^{(0)}_{\rm matter} =0~, 
\qquad  \cL^{(0)}_{\rm matter} ({\rm e}^{ {\rm i}\a}\, u^+, {\rm e}^{ -{\rm i}\a}\,  u^-)
=  \cL^{(0)}_{\rm matter} (u^+, u^-)~.
\eea
In the projective-superspace approach, couplings of the Goldstinos to $\cN=2$ supersymmetric 
matter can be described by actions of the form:
\bea
S_{\text{inter, projective}}= \frac{1}{ 2\pi}
\oint 
\frac{v_i \rd v^{i}}{ (v, w)^4}\int\rd^4 x\,({D}^-)^4 \Big\{ L^{(+4)}(v) \,\cL^{(-2)}_{\rm matter} (v) \Big\}
 ~, 
\label{5.21}
\eea
where $\cL^{(-2)}_{\rm matt}$ is a real projective superfield\footnote{See \cite{K10} for the
general definition of projective superfields.} 
of weight $-2$, 
\bea
D^+_\a \cL^{(-2)}_{\rm matter}  = {\bar D}^+_\ad \cL^{(-2)}_{\rm matter} =0~, 
\qquad \cL^{(-2)}_{\rm matter} (a v) = a^{-2} \cL^{(-2)}_{\rm matter} (v) ~.
\eea
We should point out that the interaction  $\cL^{(-2)}_{\rm matter} (v) $ 
is only required to be a holomorphic function
of $v^i$ in the vicinity of the integration contour.
Unlike the harmonic-superspace Lagrangian in (\ref{5.20}), 
the  $\cL^{(-2)}_{\rm matter} (v) $ is not expected to be globally defined over ${\mathbb C}P^1$.
The isotwistor $w_i$ in (\ref{5.21})
has to be kept constant along the integration contour.
One can show that the action (\ref{5.21}) is independent of $w_i$, see \cite{K10} for more details.
Finally, Goldstino-matter couplings can be described by actions of the form:
\bea
 S_{\text{inter, full}}=  \int \rd^4x \, \rd^4 \q \, \rd^4 {\bar \q}\, L \, \cL_{\rm matter}~, 
 \eea
 where   $\cL_{\rm matter}$ is an ordinary $\cN=2$ superfield.
 We can also generate chiral Goldstino-matter couplings using the chiral Lagrangian in (\ref{A.6}).

We would like to make a final comment. It appears that all information about the Goldstino 
superfields $\L^\pm_\a$ and $\bar \L^\pm_\ad$ is encoded in the Lagrangian $L^{(+4)}$
in the sense that these superfields can be obtained from $L^{(+4)}$ by 
applying various differential operators like $D^-_\a$, $\bar D^-_\ad$, $D^{--}$ etc. 
We believe that $L^{(+4)}$ should obey a system of constraints which involve only $L^{(+4)}$ that
completely fix its structure, 
compare with (\ref{2.27}). However, we have not been able to determine such constraints.

\section{Composite  Goldstino superfields} 
\setcounter{equation}{0}
Here we briefly discuss composite Goldstino superfields that can be used to construct 
higher-derivative Goldstino actions. 

All of the Goldstino superfields can be written as spinor covariant derivatives of  a set of ``prepotentials.'' 
Defining
\bea
 \S^{++} := \L^{\a +}\L_{\a}^+~, \qquad
\S := \L^{\a -} \L_{\a}^+~, \qquad 
  \S^{--} := \L^{\a -}\L_{\a}^-~,
 \eea
and their conjugates,  it follows that
\begin{subequations}
\bea
D_{\b}^+ \S &=& \L_{\b}^+  
 \\
D_{\b}^+ \S^{--} &=& 2 \L_{\b}^- \\
\bar{D}_{\bd}^+ \bar{ \S} &=& -  \bar{\L}_{\bd}^+ \\
\bar{D}_{\bd}^+ \bar{ \S}^{--} &=& - 2 \bar{\L}_{\bd}^- ~.
\eea
\end{subequations}
We thus can think of the scalar composites $\S$, $\S^{--}$ and their conjugates 
as  fundamental building blocks.
The rest of the $D$ algebra for $\S$  is
\begin{subequations}
\bea
\bar{D}_{\ad}^+ \S &=& 0 \\
D_{\a}^- \S &=& -\L_{\a}^- + 2\ri\, \bar{\L}^{\bd -} \L^{\b +} \partial_{\a \bd} \L_{\b}^- 
+ 2\ri\, \bar{\L}^{\bd -} \L^{\b -} \partial_{\a \bd} \L_{\b}^+ \\
\bar{D}_{\a}^- \S &=& -2\ri\, \L^{\a - } \partial_{\a \ad} \S~. 
\eea
\end{subequations}

The superfield $\bar \S^{--}$ can be viewed as  an $\cN=2$ extension
of the ${\cal N} =1$  complex linear Goldstino superfield
\cite{Kuzenko:2011ti}, eq. (\ref{2.26}), for its properties are are
\bea
-\frac14 (\bar D^+)^2
\bar{\S}^{--} &=& 1~, \qquad  ( \bar{\S}^{--})^2=0~, \qquad 
- \frac14 \,  \bar{\S}^{--} ( \bar{D}^+)^2 D_{\a}^-  \bar{\S}^{--}
=D_{\a}^- \bar{\S}^{--} ~.~~~
\eea
It is not difficult to see that $\S$ and $\bar \S$ can be expressed in terms of $\S^{--}$ 
and  $\bar \S^{--}$. Therefore, $\S^{--}$ and  $\bar \S^{--}$ contain all the information 
about the Goldstino superfields.

The superfield $\S^{++}$ is analytic, $D^+_\a \S^{++} = {\bar D}^+_\ad \S^{++} =0$, 
and obeys the covariant constancy condition
\bea
(D^{++} +\ri \, \cV^{++} ) \S^{++} = 0~, 
\qquad \cV^{++} := 2 (\L^{\a +} \pa_{\a\ad} {\bar \L}^{\ad +} +  {\bar \L}^{\ad +} \pa_{\a \ad} \L^{\a +})
=\widetilde{\cV^{++} }~.
\eea
Using $\S^{++}$,  $\bar \S^{++}$ and $\cV^{++}$, 
we can generate reduced chiral superfields of the form
\bea
W = \frac{1}{4} \int {\rm d}u \,({\bar D}^-)^2 \Big( c\, \S^{++} + {\bar c}\, \bar \S^{++} +r \, \cV^{++}
\Big)~,
\label{1.5-harmonic}
\eea
for arbitrary  complex   $c$ and real $r$ parameters. The properties of $W$ are 
\bea
{\bar D}^\ad_i W= 0~, \qquad
D^{\a i} D_\a^j W={\bar D}_\ad^{i }{\bar D}^{\ad j } \bar{W} ~.
\eea
Using such chiral superfields, we can generate higher derivative Goldstino couplings. 
\\

\noindent
{\bf Acknowledgements:}\\
We are grateful to Simon Tyler for a question leading to (\ref{2.29}).
This work is supported in part by the Australian Research Council. 

\appendix

\section{$\cN=2$ chiral construction} 
\setcounter{equation}{0}

In this appendix, we apply the technique for construction of an antichiral Golstino superfield given in section 2 to the case of spontaneously broken ${\cal N} = 2$ supersymmetry. The ${\cal N}= 2$ supersymmetry algebra is given in equation (\ref{N=2}). In the Volkov-Akulov construction, there are two Golstinos $\l_{\a i} $ (and their Hermitian conjugates $\bar{\l}_{\ad}^i$ ) associated with the broken ${\cal N} = 2$ supersymmetry generators  via the coset parametrization
\bea
g(x, \l_i(x), \bar{\l}^i(x) ) = \re^{\ri \,  ( - x^a P_a + \l^{\a}_{i} (x) Q_{\a}^i + \bar{\l}_{\ad}^i (x) \bar{Q}^{\ad}_i )} ~.
\label{A.1}
\eea
This yields the infinitesimal supersymmetry transformations
\be
\delta \l_{\a i} = \e_{\a i} - \ri \,v^{\b \bd} \partial_{\b \bd} \l_{\a i} ~,\qquad \delta \bar{\l}_{\ad}^i 
= \bar{\e}_{\ad}^i - \ri \,  v^{\b \bd} \partial_{\b \bd} \bar{\l}_{\ad}^i
\ee
with $ v^{\b \bd} = \l^{\b}_j \bar{\e}^{\bd j} - \e^{\b}_j \bar{\l}^{\bd j} .$
The analogue of the Samuel-Wess nonlinear realization, in which there is a pair of Golsdtone fields $\xi_{\a i}(x)$ which mix only with themselves under supersymmetry transformation, is based on the alternative coset parametrization
\be
g(x, \xi_i (x), \bar{\psi}^i (x) ) = \re^{\ri( - x^a P_a + \xi^{\a}_{i} (x) Q_{\a}^i)} \, \re^{ \ri  \bar{\psi}_{\ad}^i (x) \bar{Q}^{\ad}_i }~ .
\ee
This yields the supersymmetry transformations
\begin{subequations}
\bea
\delta \xi_{\a i} &=& \e_{\a} - 2 \ri \, \xi^{\b j} \bar{\e}^{\bd j} \partial_{\b \bd}  \xi_{\a i} \\
\delta \bar{\psi}_{\ad i} &=& \bar{\e}_{\ad i} -  2 \ri \, \xi^{\b j} \bar{\e}^{\bd j} \partial_{\b \bd}  \bar{\psi}_{\ad i}.
\eea
\end{subequations}
The construction of ${\cal N} = 2$ superfields associated with the Goldstinos proceeds as in the ${\cal N}=1$ case, and the resulting superfields $\Xi_{\a i}$ and $\bar{\Psi}_{\ad}^i$ satisfy the following set of constraints involving the ${\cal N} = 2$ supercovariant derivatives:
\begin{subequations}
\bea
D_{\a}^i \Xi_{\b j} &=& \e_{\b \a} \delta^i_j \\
\bar{D}_{\ad i} \Xi_{\b j} &=&  - 2 \ri \Xi_i^{\a} \partial_{\a \ad} \Xi_{\b j} \\
D_{\a}^i \bar{\Psi}_{\bd}^j &=& 0 \label{antichiral2} \\
\bar{D}_{\ad i} \bar{\Psi}_{\bd}^j &=&  - \e_{\bd \ad} \delta_i^j - 2 \ri \Xi_i^{\a } \partial_{\a \ad} \bar{\Psi}_{\bd }^j~ .
\eea
\end{subequations}
In particular, (\ref{antichiral2}) means that the superfields $\bar{\Psi}_{\ad }^i$ are antichiral, and so provide ingredients for an action obtained by integration over the antichiral subspace of ${\cal N} = 2$ superspace:
\bea 
S \propto  \int \rd^4 x \, \rd^4 {\q} \,\J^4
+\int \rd^4 x \, \rd^4 \bar{\q} \,\bar \J^4 ~, \qquad 
\label{A.6}
\eea
where $\J^4:= \frac{1}{3} \J^{ij} \J_{ij}$ and $\J^{ij}:= \J^{\a i} \J_\a^j$.
The nilpotent chiral superfield $\J^4 $ can be shown to satisfy a constraint
\bea
\J^4 \propto \J^4 {\bar D}^4 \bar \J^4~, 
\eea
which is similar to (\ref{2.27})

\begin{footnotesize}

\end{footnotesize}

\end{document}